\documentclass[10pt]{article}

\usepackage{amsmath}
\usepackage{amssymb}
\def\b{\begin{equation}}
\def\e{\end{equation}}
\def\K{{\cal K}}

\def\K{{\cal K}}

\def\b{\begin{equation}}
\def\e{\end{equation}}
\def\bd{\begin{displaystyle}}
\def\ed{\end{displaystyle}}
\def\ba{\begin{array}}
\def\ea{\end{array}}

\def\bee{\begin{enumerate}}
\def\eee{\end{enumerate}}

\def\bes{\begin{eqnarray*}}
\def\ees{\end{eqnarray*}}
\def\be{\begin{eqnarray}}
\def\ee{\end{eqnarray}}

\begin{document}
\title{A Group Theoretical Approach to Graviton Two-Point Function}
\author{S. Rahbardehghan$^1$, H. Pejhan$^2$\thanks{e-mail: h.pejhan@piau.ac.ir} and M. Elmizadeh$^2$}
\maketitle
\centerline{\it $^1$Department of Physics, Islamic Azad University, Central Branch, Tehran, Iran}
\centerline{\it $^2$Department of Physics, Science and Research Branch, Islamic Azad University, Tehran, Iran}
\vspace{1.5pt}

\begin{abstract}
Respecting the group theoretical approach, it is debated that the theory of linear conformal gravity should be formulated through a tensor field of rank-3 and mixed symmetry \cite{binegar}. Pursuing this path, such a field equation was obtained in de Sitter space \cite{takook}. In present work, considering the de Sitter ambient space notation, a proper solution to the physical part of this field equation is obtained. We have also calculated the related two-point function, which is interestingly de Sitter invariant and free of infrared divergence.
\end{abstract}

\section{Introduction}
Many people believe that conformal invariance may be the key to a future theory of quantum gravity. In this paper, we consider linear theories of gravitation, in which, not only the field equations but also the free field commutation relations are conformal invariant. The main input into this construction of linear gravity is to insist that the propagating modes must be a pair of massless particles with helicity $\pm2$. It was supposed that, a natural choice for such a field is a symmetric tensor field of rank-2. However, as proved in Ref. \cite{barut}, for the physical representation of conformal group, the value of conformal Casimir operator is $9$. While, by considering a rank-2 tensor field, related value will become $8$ \cite{binegar}. Hence, such a tensor field does not correspond to any unitary irreducible representation (UIR) of conformal group. Indeed, the mentioned physical requirement implies that the theory of linear conformal quantum gravity must be formulated in terms of a tensor field of rank-3 and mixed symmetry with conformal degree zero \cite{binegar}. Mixed symmetry means that
$$ \Psi_{abc}=-\Psi_{bac},\,\,\,\,\sum_{cycl}\Psi_{abc}=0,$$
while a field of conformal degree zero satisfies $u^d\partial_d\Psi_{abc}=0$.

On the other hand, according to the Wigner's theorem, a linear gravitational field should transform under the UIR of its space-time symmetry group. In this regard, it seems that, the theory should also be invariant under the de Sitter (dS) group as the space-time symmetry group. Our choice of dS space-time is due to the recent cosmological observations. These observational data are strongly in favour of a positive acceleration of the present universe \cite{riess}, which means, in the first approximation, our universe might currently be in a dS phase. Accordingly, a mixed symmetry tensor field of rank-3 with conformal degree zero, which transforms according to both UIRs of the conformal and de Sitter groups was achieved in Ref. \cite{takook}. In present work, a proper solution for the physical part of this conformal field equation is calculated. Then, the related conformally invariant (CI) two-point function is obtained. It is, interestingly, de Sitter invariant and free of any pathological large-distance behavior. Our method to calculate the two-point function is based on a rigorous group theoretical approach combined with a suitable adaption of Krein space quantization.

This Krein quantization method is a canonical quantization of Gupta-Bleuler type in which the Fock space is constructed over the total space ${\cal{H}}_{+} \oplus {\cal{H}}_{-}$, where $ {\cal{H}}_{+}$ ($ {\cal{H}}_{-}$) stands for the Hilbert (anti-Hilbert) space \cite{Mintchev, Bognar}. Through this construction, recently, a covariant quantization of the massless minimally coupled scalar field on de Sitter space has been carried out \cite{gazeaurenaud,De Bievre}; according to Allen's theorem \cite{Allen}, no invariant vacuum exists, therefore no covariant Hilbert space quantization is possible. It is reputed that the graviton propagator in the linear approximation on dS background suffers from the same problem. Actually, for largely separated points, it has a pathological behavior (infrared divergence) and also breaks the de Sitter invariance \cite{AllenT, floratos, antoniadis}.\footnote{On this basis, it has been proposed that infrared divergence might lead to instability of de Sitter space \cite{ford, antoni1}. So, some authors, by considering a dS field operator for linear gravity in terms of flat coordinates (it covers only one-half of the de Sitter hyperboloid), have investigated the possibility of quantum instability and have found a quantum field, which violates the de Sitter invariance \cite{tsamis}. However, recently it was shown that the infrared divergence of the graviton propagator in the one-loop approximation is gauge dependent, therefore, it should not appear in an effective way as a physical quantity \cite{antoni,higuchi,vega}.} Respecting Krein quantization method, however, these difficulties are solved. The singularity of the Wightman two-point function, which appears due to the zero mode problem of the Laplace-Beltrami operator on dS space \cite{Allen}, is removed, and interestingly the de Sitter invariance is survived. [To achieve a detailed construction of the quantization method, the unitarity condition and compatibility with (Hilbert space) QFT's counterpart in the Minkowskian limit, one could refer to Refs. \cite{Garidi,Pejhan}.]

The layout of the paper is as follows. In section (2), we briefly introduce the notations, and in particular, study the CI massless spin-2 wave equations in dS space. In Section (3), by focusing on the physical part of the field equations, the corresponding solution is calculated. It is actually constructed over the massless minimally coupled scalar field. In section (4), we calculate the two-point function ${\cal W}_{\alpha\beta\gamma\alpha'\beta'\gamma'}(x,x')$ in the ambient space notations. Especially, it is shown that, through Krein space quantization, we are capable of calculating the physical graviton two-point function, that is dS-invariant and free of any divergences. Finally, in section (5), the results of the paper are discussed. Some mathematical relations are given in the appendices.

\setcounter{equation}{0}
\section{De Sitter Space and Dirac's Six-Cone Formalism}
\subsection{De Sitter space}
The de Sitter solution to the cosmological Einstein field equation (with positive cosmological constant $\Lambda$) can be viewed as a one-sheeted hyperboloid embedded in a five dimensional Minkowski space $M^5$
\begin{equation} {X_H}= \{ x \in {R}^5 ; x^2={\eta}_{\alpha\beta} {x^{\alpha}} {x^\beta} = -H^{-2}= -\frac{3}{\Lambda} \}, \;\;\alpha,\beta= 0,1,2,3,4, \end{equation}
where $\eta_{\alpha\beta}=$ diag$(1,-1,-1,-1,-1)$ and $H$ is the Hubble parameter. The dS metric is
$$ ds^2=\eta_{\alpha\beta}dx^{\alpha}dx^{\beta}=g_{\mu\nu}^{dS}dX^{\mu}dX^{\nu},\;\;\mu,\nu=0,1,2,3.$$
We use $x^\alpha$ for ambient space formalism (five global coordinates) whereas $X^\mu$ stand for de Sitter intrinsic coordinates (four local coordinates). In what follows, the ambient space notation is used, because working in the embedding space has two advantages, first it is close to the group theoretical language and second the equations are obtained in an easer way than they might be found in de Sitter intrinsic space.

The dS kinematical group is the $10$-parameter group $SO_0(1,4)$ (connected component of the identity in $O(1,4)$), for which there are two Casimir operators
\begin{equation} \label{2casimir} Q^{(1)}=-\frac{1}{2}L^{\alpha\beta}L_{\alpha\beta},\;\;\;\;\ Q^{(2)}=-W_{\alpha}W^{\alpha},\end{equation}
where $W_{\alpha}=-\frac{1}{8}\epsilon_{\alpha\beta\gamma\sigma\eta}L^{\beta\gamma}L^{\sigma\eta}$ and $\epsilon_{\alpha\beta\gamma\sigma\eta}$ is the antisymmetric tensor in the ambient space notation with $\epsilon_{01234}=1$. The generator of the de Sitter group is $L_{\alpha\beta}=M_{\alpha\beta}+{{\sum}_{\alpha\beta}}$, in which, the action of the orbital, $M_{\alpha\beta}$, and the spinorial, ${\sum}_{\alpha\beta}$, parts are respectively defined by \cite{J.P. Gazeau}
\begin{equation} \label{S and M}\begin{aligned}
M_{\alpha\beta}\equiv&-i(x_{\alpha}\partial_{\beta}-x_{\beta}\partial_{\alpha})= -i(x_{\alpha}\bar\partial_{\beta}-x_{\beta}\bar\partial_{\alpha}),\\
{\sum}_{\alpha\beta}\K_{\gamma\delta ...}\equiv&-i(\eta_{\alpha\gamma}\K_{\beta\delta ...}-\eta_{\beta\gamma} \K_{\alpha\delta ...} + \eta_{\alpha\delta}\K_{\gamma\beta ...}-\eta_{\beta\delta}  \K_{\gamma\alpha ...} + ...).
\end{aligned}\end{equation}
$\bar\partial_{\alpha}$ is the tangential (or transverse) derivative on dS space, defined by
\begin{equation} \label{transverse} \bar\partial_{\alpha}=\theta_{\alpha\beta}\partial^{\beta}=\partial_{\alpha}+H^2x_{\alpha}x\cdot\partial,\,\,\,\mbox{with}\,\,\,x\cdot\bar\partial=0\,,\end{equation}
and $\theta_{\alpha\beta}$ is the transverse projector ($\theta_{\alpha\beta} = \eta_{\alpha\beta} + H^2 x_\alpha x_\beta$).

The operator $Q^{(1)}$ commutes with the action of the group generators, thus, it is constant in each UIR. The eigenvalues of $Q^{(1)}$ can be used to classify the UIRs, i.e.
\begin{equation} (Q^{(1)}-\langle Q^{(1)}\rangle){\cal K}(x)=0. \end{equation}
Following Dixmier \cite{dixmier}, one can get a classification scheme considering a pair $(p,q)$ of parameters involved in the following possible spectral values of the Casimir operators
\begin{equation} Q^{(1)}=\left(-p(p+1)-(q+1)(q-2)\right)I_d ,\qquad\quad Q^{(2)}=\left(-p(p+1)q(q-1)\right)I_d\,.\end{equation}
According to the range of values of the parameters $p$ and $q$, there exist three distinct types of UIRs for $SO(1,4)$ \cite{dixmier, takahashi}, namely: principal, complementary and discrete series. In the case of the principal and complementary series, the flat limit compels the value of $p$ to bear the meaning of spin. For the discrete series case, the only representation which has a physically meaningful Minkowskian counterpart is $p=q$ case. For more mathematical details of the group contraction and the physical principles underlying the relationship between dS and Poincar\'e groups, one can refer to Refs. \cite{levy, bacry}.

The spin-2 tensor representations relevant to the present work are as follows:\\
I) The UIRs $U^{2,\nu}$ in the principal series, $p = s = 2$ and $q = \frac{1}{2} + i\nu$, correspond to the Casimir spectral values
\begin{equation} \langle Q^{\nu}\rangle = {\nu}^2 - \frac{15}{4}, \;\;\; {\nu}\in {R}, \end{equation}
in which $U^{2,\nu}$ and $U^{2,-\nu}$ are equivalent.\\
II) The UIRs $V^{2,q}$ in the complementary series, $p = s = 2$ and $ q - q^2 = \mu $, correspond to
\begin{equation} \langle Q^{\mu} \rangle = q - q^2 -4 \equiv{ {\mu} -4}, \;\;\; 0 < \mu < \frac{1}{4}. \end{equation}
III) The UIRs $\Pi^{\pm}_{2,q}$ in the discrete series, $p = s = 2$, correspond to
\begin{equation} \label{2.12} \langle Q^{(1)}\rangle = -4, \; q=1 \; (\Pi^{\pm}_{2,1}); \;\; \langle Q^{(2)}\rangle = -6, \; q=2 \;(\Pi^{\pm}_{2,2}).\end{equation}
Regarding the de Sitter group, the "massless"\footnote{It should be noted that in de Sitter space, the mass concept does not exist by itself as a conserved quantity. It is actually referred to the conformal invariance (propagating on the dS light cone). The term "massive", however, is used in reference to fields that in the flat limit would be reduced to massive Minkowskian fields \cite{barut}.} spin-2 field is symbolized by $\Pi^{\pm}_{2,2}$ and $\Pi^{\pm}_{2,1}$ (the signs $\pm$ correspond to the two types of helicity for the massless tensor field). In these cases, the two representations $\Pi^{\pm}_{2,2}$, in the discrete series with $ p = q = 2 $, have a Minkowskian interpretation. It is worth to mention that $p$ and $q$ do not bear the meaning of mass and spin. For discrete series in the limit $H \rightarrow 0$, $p = q = s$ are veritably none other than spin.

The compact subgroup of the conformal group $SO(2,4)$ is $SO(2)\otimes SO(4)$, in which, by considering $E$ as the eigenvalues of the conformal energy generator of $SO(2)$ and $(j_1,j_2)$ as the $(2j_1+1)(2j_2+1)$ dimensional representation of $SO(4)=SU(2)\otimes SU(2)$, the mathematical symbols $C(E;j_1,j_2)$ can be used to denote the irreducible projective representation of the conformal group. The representation $\Pi^+_{2,2}$ has a unique extension to a direct sum of two UIRs $C(3;2,0)$ and $C(-3;2,0)$ of the conformal group, with positive and negative energies respectively \cite{barut,levy}. The latter is restricted to the massless Poincar\'e UIRs $P^>(0, 2)$ and $P^<(0,2)$ with positive and negative energies respectively. $ {\cal P}^{ \stackrel{>} {<}}(0,2)$ (resp. $ {\cal P}^{\stackrel{>}{<}}(0,-2)$) are the massless Poincar\'e UIRs with positive and negative energies and positive (resp. negative) helicity. The following diagrams elucidate these connections
\begin{equation}
\left.
\begin{array}{ccccccc}
&& {\cal C}(3,2,0)& &{\cal C}(3,2,0)&\hookleftarrow &{\cal P}^{>}(0,2)\\
\Pi^+_{2,2} &\hookrightarrow &\oplus&\stackrel{H=0}{\longrightarrow} & \oplus  & &\oplus\\
&& {\cal C}(-3,2,0)& & {\cal C}(-3,2,0) &\hookleftarrow &{\cal
P}^{<}(0,2),\\
\end{array}
\right.
\end{equation}

\begin{equation}
\left.
\begin{array}{ccccccc}
&& {\cal C}(3,0,2)& &{\cal C}(3,0,2)&\hookleftarrow &{\cal P}^{>}(0,-2)\\
\Pi^-_{2,2} &\hookrightarrow &\oplus&\stackrel{H=0}{\longrightarrow}&\oplus &&\oplus\\
&& {\cal C}(-3,0,2)&& {\cal C}(-3,0,2)&\hookleftarrow &{\cal P}^{<}(0,-2),\\
\end{array}
\right.
\end{equation}
the arrows $\hookrightarrow $ indicate unique extension. It is worth to mention that the representations $\Pi^{\pm}_{2,1}$ do not have corresponding zero curvature limit \cite{levy, bacry}.

\subsection{Dirac's six cone formalism and conformal-invariant field equations}
The concept of conformal space and six-cone formalism was firstly used by Dirac to obtain the field equations for spinor and vector fields in $1 + 3$ dimensional space-time in the conformally covariant form \cite{dirac}. He suggested a manifestly CI formulation in which the Minkowski coordinates are embedded as the hypersurface $\eta_{ab}u^a u^b =0, \; (a,b=0,1,2,3,4,5), \; \eta_{ab}=\mbox{diag}(1,-1,-1,-1,-1,1) $ in $R^6$. Then the fields are extended by homogeneity requirements to the whole of the space of homogeneous coordinates, namely $R^6$. Reduction to four dimensions (physical space-time) is carried out by projection, that is, by fixing the degrees of homogeneity of all fields. Wave equations, subsidiary conditions, etc., must be expressed in terms of well-defined operators which are determined intrinsically on the cone (they actually map tensor fields to tensor fields with the same rank on the cone $u^2=0$). Thus, the obtained equations through this method are conformally invariant. This approach to the conformal symmetry was then developed by Mack and Salam \cite{mack} and many others \cite{kastrup}.

Considering this method in de Sitter space provides us with the opportunity to acquire the CI field equations for massless scalar, vector and tensor fields \cite{takook,dehghani,behroozi}. It has been shown that these CI equations in the zero curvature limit $(H \rightarrow 0)$ would be reduced exactly to their Minkowskian counterparts, e.g., Maxwell equations are achieved from the vector field case \cite{dehghani,behroozi}.

As discussed in section (1), we are interested in the conformal invariance properties of massless spin-2 field in dS space, i.e. the dS linear gravity. Generalizing the group theoretical approach, based on what was proposed by Binegar et al \cite{binegar} to de Sitter space and using a mixed symmetry tensor field of rank-3 with conformal degree zero, the related CI wave equation in dS space is best obtained as follows  \cite{takook}\footnote{Note: for sake of simplicity, from now on, we take $H = 1$ and use the notation $ \bar\partial^\alpha F_{\alpha\beta\gamma} \equiv \bar\partial\cdot F_{\cdot\beta\gamma} $}
\begin{equation}\begin{aligned} \label{ci}
2 Q_{0}^{(1)}(Q_{0}^{(1)} -2)(F_{\alpha\beta\gamma} - \frac{1}{4} x_\gamma {\cal{A}}_{\alpha\beta}) + (\bar\partial_\alpha + 3x_\alpha)(Q_{0}^{(1)} - 2)(4\bar\partial\cdot F_{\cdot\beta\gamma}- {\cal{A}}_{\gamma\beta} -x_\gamma \bar\partial\cdot {\cal{A}}_{\cdot\beta})\\
+(\bar\partial_\beta + 3x_\beta)(Q_{0}^{(1)} - 2)(4\bar\partial\cdot F_{\alpha\cdot\gamma}- {\cal{A}}_{\alpha\gamma} -x_\gamma \bar\partial\cdot {\cal{A}}_{\cdot\alpha})=0,
\end{aligned}\end{equation}
in which $ Q_{0}^{(1)} = -\frac{1}{2} M^{\alpha\beta} M_{\alpha\beta}$, $F_{\alpha\beta\gamma}$ is the projected tensor field to dS space and $ {\cal{A}}_{\alpha\beta}\equiv \bar\partial^\gamma F_{\alpha\beta\gamma} - x_\alpha F^\gamma _{\gamma\beta} + x_\beta F^\gamma_{\gamma\alpha}.$
Now, by imposing the mixed symmetry, transversality, divergenceless and traceless conditions on the tensor field $F_{\alpha\beta\gamma}$, which are
necessary for UIRs of the dS and conformal groups, the CI equation (\ref{ci}) reduces to (see Appendix A)
\begin{equation}\label{2.14-1} Q_{0}^{(1)} (Q_{0}^{(1)} - 2)F_{\alpha\beta\gamma}=0, \;\; \mbox{or equivalently,}\;\; (Q^{(1)} + 6)(Q^{(1)} + 4)F_{\alpha\beta\gamma}=0.\end{equation}
Obviously this CI field corresponds to the two representations of discrete series, $\Pi^{\pm}_{2,1}$ and $\Pi^{\pm}_{2,2}$ (the physical representation of the de Sitter group). Accordingly, the parameter $p$ does have a physical significance. It is indeed spin. In what follows, however, we are only interested in the tensor field that corresponds to the representations of $\Pi^{\pm}_{2,2}$, i.e.
\begin{equation} \label{base} (Q^{(1)} + 6)F_{\alpha\beta\gamma}=0.\end{equation}
As already pointed, these are actually the only two representations in the discrete series which have a Minkowskian interpretation.

\setcounter{equation}{0}
\section{De Sitter Field Solution}
In this section, we want to obtain solution of the physical part of the CI field equation. To start, we consider the most generic form of $ F_{\alpha\beta\gamma}$ as follows
\begin{equation} \label{F_} F_{\alpha\beta\gamma}= ({\bar\partial}_\alpha + x_\alpha)K_{\beta\gamma} - ({\bar\partial}_\beta + x_\beta)K_{\alpha\gamma} + \bar Z_\alpha H_{\beta\gamma} - \bar Z_\beta H_{\alpha\gamma},\end{equation}
where $ K_{\alpha\beta} $ and $ H_{\alpha\beta} $ are two rank-2 tensor fields and $ Z $ is a 5-dimensional constant vector. Bar over the vector makes it a tangential (or transverse) vector on dS space (see (\ref{transverse})). Imposing the mixed symmetry, transversality, divergenceless and traceless conditions on $ F_{\alpha\beta\gamma}$, which are needed in order to relate it to the physical representation, leads to
\begin{equation} \begin{aligned} \label{phi1}
K_{\alpha\beta} &= K_{\beta\alpha},\hspace{3mm} x\cdot K_{\cdot\beta}= x\cdot K_{\alpha\cdot}=0,\\
H_{\alpha\beta}=H_{\beta\alpha},\hspace{3mm} x\cdot H_{\cdot\beta}&= x\cdot H_{\alpha\cdot}=0, \hspace{3mm} \bar\partial\cdot H_{\cdot\beta}= \bar\partial\cdot H_{\alpha\cdot}=0, \hspace{3mm} {\cal{H}'}=0,
\end{aligned} \end{equation}
where ${\cal{H}'} = H_\alpha^\alpha$ is the trace of $H_{\alpha\beta}$. In addition one obtains useful relations as follows
\begin{equation} \label{conditions} \begin{aligned}
(Q_{0}^{(1)} -2)K_{\beta\gamma} + ({\bar\partial}_\beta + 2x_\beta)\bar\partial\cdot K_{\cdot\gamma} - (Z\cdot\bar\partial + 3x\cdot Z)H_{\beta\gamma} + x_\beta Z\cdot H_{\cdot\gamma} =&0,\;\;\;\;\; (I)\\
(\bar\partial_\alpha + 2x_\alpha)\bar\partial\cdot K_{\cdot\beta} - (\bar\partial_\beta + 2x_\beta)\bar\partial\cdot K_{\alpha\cdot} + x_\alpha Z\cdot H_{\cdot\beta} - x_\beta Z\cdot H_{\alpha\cdot} =&0,\;\;\;\; (II) \\
(\bar\partial_\alpha + x_\alpha){\cal{K}}' - \bar\partial\cdot K_{\alpha\cdot} - Z\cdot H_{\alpha\cdot} =& 0,\;\;\; (III)
\end{aligned} \end{equation}
${\cal{K}'} = K_\alpha^\alpha$ is the trace of $K_{\alpha\beta}$.

On the other hand, substituting $ F_{\alpha\beta\gamma}$ in (\ref{base}), results in [From now on, in order to get shorthand equations, we define a symmetrizer operator, i.e. $ S_{\alpha\beta} K_{\alpha\beta} \equiv K_{\alpha\beta} + K_{\beta\alpha}$, and an anti-symmetrizer operator, i.e. $ \bar S_{\alpha\beta} K_{\alpha\beta} \equiv K_{\alpha\beta} - K_{\beta\alpha}$.]
\begin{equation}\label{ar}
\left\{\begin{array}{rl} \bar S_{\alpha\beta}\Big( (\bar\partial_\alpha + 3x_\alpha)Q_{0}^{(1)} - 4x_\alpha \Big)K_{\beta\gamma}=&
\bar S_{\alpha\beta}\Big( (8x_\alpha + 2\bar\partial_\alpha)(x\cdot Z) + 2x_\alpha (Z\cdot\bar\partial) \Big)H_{\beta\gamma} ,\,\,\,\,\,\,
{(I)}\vspace{2mm}\\\vspace{2mm}
Q_{0}^{(1)} H_{\beta\gamma} =&0. \hspace{6.6cm} {(II)}\\
\end{array}\right.
\end{equation}

From the Eq. (\ref{ar}-I) along with the conditions given in (\ref{phi1}), (\ref{conditions}) and after following the procedure given in Appendix B, it is proved that $ K_{\beta\gamma} $ can be written in terms of $H_{\beta\gamma}$ as
\begin{equation} \label{kh} K_{\beta\gamma}(x)= \Big( -\frac{1}{2}(x\cdot Z) + \frac{1}{8} (Z\cdot\bar\partial) \Big) H_{\beta\gamma} - \frac{1}{8} \Big( x_\beta Z\cdot H_{\cdot\gamma} + x_\gamma Z\cdot H_{\beta\cdot} \Big).\end{equation}
Thus we can construct the tensor field (\ref{F_}) as follows
\begin{equation} \label{construct} \begin{aligned}
F_{\alpha\beta\gamma}(x)= \bar S_{\alpha\beta} \Big[ ({\bar\partial}_\alpha + x_\alpha) \Big( -\frac{1}{2}(x\cdot Z)& + \frac{1}{8} (Z\cdot\bar\partial) \Big) + \bar Z_\alpha \Big] H_{\beta\gamma}\\& - \frac{1}{8} \bar S_{\alpha\beta} ({\bar\partial}_\alpha + x_\alpha) \Big( x_\beta Z\cdot H_{\cdot\gamma} + x_\gamma Z\cdot H_{\beta\cdot} \Big),
\end{aligned} \end{equation}
where $ H_{\beta\gamma} $ must satisfy the Eq. (\ref{ar}-$II$). After utilizing the similar procedure, which is given in Ref. \cite{dehghani}, it is proved that
\begin{equation} \label{equat} \begin{aligned}
H(x)= \Big[-\frac{2}{3} & \theta Z_1\cdot + {S}\bar Z_{1}\\& + \frac{1}{3} S(\bar\partial - x) \Big( \frac{1}{9} \bar\partial Z_1\cdot + x\cdot Z_1 \Big) \Big] \Big[ \bar Z_{2} - \frac{1}{2} \bar\partial \Big( Z_2 \cdot\bar\partial + 2 x\cdot Z_2\Big) \Big]\phi,
\end{aligned} \end{equation}
where $Z_1,\;Z_2$ and $Z_3$ are another 5-dimensional constant vectors and $\phi$ is the massless minimally coupled scalar field.

\setcounter{equation}{0}
\section{Two-Point Function}
In this section, we deal with conformally invariant two-point function of the massless spin-2 field. We write the two-point function in dS space in terms of bi-tensors which are called maximally symmetric if they respect dS invariance. Bi-tensors are functions of two points $(x, x')$ and behave like tensors under coordinate transformations at each points \cite{allen2}. Moreover, the dS axiomatic field theory is constructed over bi-tensor Wightman two-point function \cite{bros2, garidigarou}. On this basis, the two-point function is given by
\begin{equation} \label{b:} {\cal W}_{\alpha\beta\gamma \alpha'\beta'\gamma'}(x,x')=\langle \Omega| F_{\alpha\beta\gamma}(x)F_{\alpha'\beta'\gamma'}(x')|\Omega  \rangle ,\end{equation}
where $x,x'\in X_H$ and $|\Omega\rangle $ is the Fock-vacuum state. In this respect, by considering the Eqs. (\ref{F_}) and (\ref{b:}), the following form for two-point function is proposed\footnote{Note that, the primed operators act only on the primed coordinates and vise versa, so that $\bar\partial \bar\partial'=\bar\partial' \bar\partial$.}
\begin{equation} \label{4.2} {\cal W}_{\alpha\beta\gamma\alpha'\beta'\gamma'}(x,x')= {\bar{S}}_{\alpha\beta} ({\bar\partial}_\alpha + x_\alpha) \Big( {\bar{S'}}_{\alpha'\beta'} ({{\bar\partial}'}_{{\alpha}'} + {x'}_{{\alpha}'}) {\cal{W}}^K_{\beta\gamma\beta'\gamma'} (x,x') \Big)+ {\bar{S}}_{\alpha\beta} {\bar{S'}}_{\alpha'\beta'} \Big( (\theta_\alpha\cdot{\theta'}_{\alpha'}){\cal{W}}^H_{\beta\gamma\beta'\gamma'} (x,x') \Big).\end{equation}
${\cal{W}}^K_{\beta\gamma\beta'\gamma'}$ and ${\cal{W}}^H_{\beta\gamma\beta'\gamma'}$ are two transverse bi-tensor two-point functions which will be determined through the similar procedure of the previous section. Actually, the two-point function (\ref{4.2}) must verify the Eq. (\ref{base}) with respect to $x$ and $x'$ (without any difference), and also the physical requirements; mixed symmetry, transversality, divergenceless and traceless conditions, which imply that
\begin{itemize}
\item ${\cal W}_{\alpha\beta\gamma\alpha'\beta'\gamma'} = - {\cal W}_{\beta\alpha\gamma\alpha'\beta'\gamma'}$,  $\;\;{\cal W}_{\alpha\beta\gamma\alpha'\beta'\gamma'} = - {\cal W}_{\alpha\beta\gamma\beta'\alpha'\gamma'}$.
\item $ \sum_{cycl\{ \alpha,\beta,\gamma\}}{\cal W}_{\alpha\beta\gamma\alpha'\beta'\gamma'}=0$,  $\;\;\sum_{cycl\{\alpha',\beta',\gamma'\}}{\cal W}_{\alpha\beta\gamma\alpha'\beta'\gamma'}=0$.

\item $ x\cdot {\cal W}_{\cdot\beta\gamma\alpha'\beta'\gamma'} = ... =0 $,  $\;\; x'\cdot {\cal W}_{\alpha\beta\gamma\cdot\beta'\gamma'}= ... =0$.

\item $\bar\partial\cdot {\cal W}_{\cdot\beta\gamma\alpha'\beta'\gamma'}= ... =0 $,  $\;\;\bar\partial'\cdot {\cal W}_{\alpha\beta\gamma\cdot\beta'\gamma'}= ... =0 $.

\item ${\cal W}^{\beta}_{\alpha\beta\alpha'\beta'\gamma'}=0$,  $\;\;{\cal W}^{\beta'}_{\alpha\beta\gamma\alpha'\beta'}=0$.
\end{itemize}

At the first step, with regard to the above considerations, we investigate the two-point function (\ref{4.2}) with the choice of $x$. Accordingly, by imposing the mentioned requirements on the two-point function, we have
\begin{equation}\label{I} \begin{aligned}
{\cal{W}}^K_{\alpha\beta\alpha'\beta'}={\cal{W}}^K_{\beta\alpha\alpha'\beta'},\;\; x\cdot{\cal{W}}^K_{\cdot\beta\alpha'\beta'}=x\cdot{\cal{W}}^K_{\alpha\cdot\alpha'\beta'}=0,\\
{\cal{W}}^H_{\alpha\beta\alpha'\beta'}={\cal{W}}^H_{\beta\alpha\alpha'\beta'},\;\; x\cdot{\cal{W}}^H_{\cdot\beta\alpha'\beta'}=x\cdot{\cal{W}}^H_{\alpha\cdot\alpha'\beta'}=0,\\
\bar\partial\cdot{\cal{W}}^H_{\cdot\beta\alpha'\beta'}=\bar\partial\cdot{\cal{W}}^H_{\alpha\cdot\alpha'\beta'}=0,\;\; {{\cal{W}}^{H}}^\alpha_{\alpha\alpha'\beta'}=0,\;\;\;\;\;
\end{aligned} \end{equation}
and also
\begin{equation} \label{II}\begin{aligned}
(Q_{0}^{(1)} -2)  \Big( {\bar{S'}}_{\alpha'\beta'} ({{\bar\partial}'}_{{\alpha}'} + {x'}_{{\alpha}'}) {\cal{W}}^K_{\beta\gamma\beta'\gamma'} \Big) + ({\bar\partial}_\beta + 2x_\beta)\Big( {\bar{S'}}_{\alpha'\beta'} ({{\bar\partial}'}_{{\alpha}'} + {x'}_{{\alpha}'}) \bar\partial\cdot{\cal{W}}^K_{\cdot\gamma\beta'\gamma'} \Big)&\\
- {\bar{S'}}_{\alpha'\beta'}(\theta'_{\alpha'}\cdot\bar\partial+ 3x\cdot \theta'_{\alpha'}){\cal{W}}^H_{\beta\gamma\beta'\gamma'} + {\bar{S'}}_{\alpha'\beta'} x_\beta \theta'_{\alpha'}\cdot {\cal{W}}^H_{\cdot\gamma\beta'\gamma'} &=0,\\
{\bar{S}}_{\alpha\beta}(\bar\partial_\alpha + 2x_\alpha) \Big( {\bar{S'}}_{\alpha'\beta'} ({{\bar\partial}'}_{{\alpha}'} + {x'}_{{\alpha}'}) \bar\partial\cdot{\cal{W}}^K_{\cdot\beta\beta'\gamma'}\Big) + {\bar{S}}_{\alpha\beta} x_\alpha \Big( {\bar{S'}}_{\alpha'\beta'} \theta'_{\alpha'}\cdot{\cal{W}}^H_{\cdot\beta\beta'\gamma'}\Big) &=0,\\
(\bar\partial_\alpha + x_\alpha)\Big({\bar{S'}}_{\alpha'\beta'} ({{\bar\partial}'}_{{\alpha}'} + {x'}_{{\alpha}'}){{\cal{W}}^K}^\beta_{\beta\beta'\gamma'} \Big) - {\bar{S'}}_{\alpha'\beta'}({{\bar\partial}'}_{{\alpha}'} + {x'}_{{\alpha}'}) \bar\partial\cdot {\cal{W}}^K_{\alpha\cdot\beta'\gamma'}- {\bar{S'}}_{\alpha'\beta'} \theta'_{\alpha'}\cdot{\cal{W}}^H_{\alpha\cdot\beta'\gamma'} &=0.
\end{aligned} \end{equation}
On the other side, ${\cal W}_{\alpha\beta\gamma \alpha'\beta'\gamma'}(x,x')$ must satisfy the Eq. $(\ref{base})$, so one can easily show
\begin{equation}\label{IIIa} \bar S_{\alpha\beta}{\bar{S'}}_{\alpha'\beta'} ({{\bar\partial}'}_{{\alpha}'} + {x'}_{{\alpha}'})\Big( (\bar\partial_\alpha + 3x_\alpha)Q_{0}^{(1)} - 4x_\alpha \Big){\cal{W}}^K_{\beta\gamma\beta'\gamma'} =\bar S_{\alpha\beta}{\bar{S'}}_{\alpha'\beta'} \Big( (8x_\alpha + 2\bar\partial_\alpha)(x\cdot \theta'_{\alpha'}) + 2x_\alpha(\theta'_{\alpha'}\cdot\bar\partial)\Big){\cal{W}}^H_{\beta\gamma\beta'\gamma'},\end{equation}
\begin{equation}\label{IIIb} Q_{0}^{(1)} {\cal{W}}^H_{\beta\gamma\beta'\gamma'} =0.\;\;\;\;\;\;\;\;\;\;\;\;\;\;\;\;\;\;\end{equation}

Consistently with (\ref{I}), (\ref{II}), (\ref{IIIa}) and based on the procedure presented in section (3), it is a matter of simple calculation to get
\begin{equation} \label{W^K} \begin{aligned}
{\bar{S'}}_{\alpha'\beta'} ({{\bar\partial}'}_{{\alpha}'} + {x'}_{{\alpha}'}) {\cal{W}}^K_{\beta\gamma\beta'\gamma'}(x,x') =& \; {\bar{S'}}_{\alpha'\beta'} \Big( -\frac{1}{2}(x\cdot{\theta'}_{\alpha'})+ \frac{1}{8} ({\theta'}_{\alpha'}\cdot \bar\partial)\Big){\cal{W}}^H_{\beta\gamma\beta'\gamma'} (x,x')\\
- \frac{1}{8} {\bar{S'}}_{\alpha'\beta'} &\Big( x_\beta {\theta'}_{\alpha'}\cdot {\cal{W}}^H_{\cdot\gamma\beta'\gamma'} (x,x') + x_\gamma {\theta'}_{\alpha'}\cdot {\cal{W}}^H_{\beta\cdot\beta'\gamma'} (x,x')\Big).
\end{aligned} \end{equation}
Then, according to the Eqs. (\ref{4.2}) and (\ref{W^K}), we have
\begin{equation} \label{wituout} \begin{aligned}
{\cal W}_{\alpha\beta\gamma\alpha'\beta'\gamma'}(x,x')&={\bar{S}}_{\alpha \beta}{\bar{S}'}_{\alpha' \beta'}\Big[ (\bar\partial_\alpha + x_\alpha) \Big( -\frac{1}{2} (x\cdot \theta'_{\alpha'}) + \frac{1}{8} (\theta'_{\alpha'}\cdot\bar\partial) \Big) + (\theta _{\alpha}\cdot \theta'_{\alpha'})\Big]{\cal{W}^H_{\beta\gamma\beta'\gamma'}}(x,x') \\ &-\frac{1}{8} {\bar{S}}_{\alpha \beta}{\bar{S}'}_{\alpha' \beta'} (\bar\partial_\alpha + x_\alpha) \Big(x_\beta \theta'_{\alpha'}\cdot {\cal{W}^H_{\cdot\gamma\beta'\gamma'}}(x,x') + x_\gamma\theta'_{\alpha'}\cdot {\cal{W}^H_{\beta\cdot\beta'\gamma'}}(x,x')\Big),
\end{aligned} \end{equation}
here $ {\cal{W}^H_{\beta\gamma\beta'\gamma'}}(x,x') $ applies in the Eq. (\ref{IIIb}). Meanwhile, such transverse function was found in Ref. \cite{dehghani} as
\begin{equation} \begin{aligned} \label{4.13}
{\cal{W}^H}(x,x')= \Big(-\frac{2}{3} S'\theta\theta'\cdot + SS'\theta\cdot\theta' &\\ + \frac{1}{3}SS'(\bar\partial - x)[x\cdot\theta'& + \frac{1}{9}\bar\partial\theta'\cdot] \Big) \Big(\theta\cdot\theta' - \frac{1}{2}\bar\partial[\theta'\cdot\bar\partial + 2\theta'\cdot x] \Big){\cal W}_{mc}(x,x'),
\end{aligned} \end{equation}
${\cal W}_{mc} $ is the two-point function for the minimally coupled massless scalar field in dS space.

Now, at the second step, we investigate the two-point function (\ref{4.2}) with respect to $x'$. In this case, the physical requirements imply that
$$ {\cal{W}}^{\{K,H\}}_{\alpha\beta\alpha'\beta'}={\cal{W}}^{\{K,H\}}_{\alpha\beta\beta'\alpha'},\;\; x'\cdot{\cal{W}}^{\{K,H\}}_{\alpha\beta\cdot\beta'}=x'\cdot{\cal{W}}^{\{K,H\}}_{\alpha\beta\alpha'\cdot}=0,\;\; \bar\partial'\cdot{\cal{W}}^H_{\alpha\beta\cdot\beta'}=\bar\partial'\cdot{\cal{W}}^H_{\alpha\beta\alpha'\cdot}=0,\;\; {{\cal{W}}^{H}}^{\alpha'}_{\alpha\beta\alpha'}=0,$$
in addition
\begin{equation}\begin{aligned}
({Q'}_{0}^{(1)} -2)\Big( {\bar{S}}_{\alpha\beta} ({{\bar\partial}}_{{\alpha}} + {x}_{{\alpha}}) {\cal{W}}^K_{\beta\gamma\beta'\gamma'}\Big)+
({\bar\partial'}_{\beta'} + 2x'_{\beta'})\Big( {\bar{S}}_{\alpha\beta} ({{\bar\partial}}_{{\alpha}} + {x}_{{\alpha}}) \bar\partial'\cdot{\cal{W}}^K_{\beta\gamma\cdot\gamma'} \Big)\\
- {\bar{S}}_{\alpha\beta}(\theta_{\alpha}\cdot\bar\partial' + 3x'\cdot\theta_{\alpha}){\cal{W}}^H_{\beta\gamma\beta'\gamma'} + {\bar{S}}_{\alpha\beta} {x'}_{\beta'} \theta_{\alpha}\cdot {\cal{W}}^H_{\beta\gamma\cdot\gamma'} =&0,\\
{\bar{S'}}_{\alpha'\beta'}(\bar{\partial'}_{\alpha'} + 2{x'}_{\alpha'}) \Big( {\bar{S}}_{\alpha\beta} ({{\bar\partial}}_{{\alpha}} + {x}_{{\alpha}}) \bar\partial'\cdot{\cal{W}}^K_{\beta\gamma\cdot\beta'}\Big) + {\bar{S'}}_{\alpha'\beta'} {x'}_{\alpha'} \Big({\bar{S}}_{\alpha\beta}\theta_{\alpha}\cdot{\cal{W}}^H_{\beta\gamma\cdot\beta'}\Big) =&0, \\
(\bar\partial'_{\alpha'} + {x'}_{\alpha'})\Big({\bar{S}}_{\alpha\beta} ({{\bar\partial}}_{{\alpha}} + {x}_{{\alpha}}){{\cal{W}}^K}^{\beta'}_{\beta\gamma\beta'} \Big) - {\bar{S}}_{\alpha\beta}({{\bar\partial}}_{{\alpha}} + {x}_{{\alpha}}) \bar\partial'\cdot {\cal{W}}^K_{\beta\gamma\alpha'\cdot}- {\bar{S}}_{\alpha\beta} \theta_{\alpha}\cdot{\cal{W}}^H_{\beta\gamma\alpha'\cdot} =& 0.
\end{aligned}\end{equation}
Substituting ${\cal W}_{\alpha\beta\gamma \alpha'\beta'\gamma'}(x,x')$ into the Eq. (\ref{base}) leads to
$$ \bar S_{\alpha\beta}{\bar{S'}}_{\alpha'\beta'} ({{\bar\partial}}_{{\alpha}} + {x}_{{\alpha}})\Big( (\bar\partial'_{\alpha'} + 3{x'}_{\alpha'}){Q'}_{0}^{(1)} - 4{x'}_{\alpha'} \Big){\cal{W}}^K_{\beta\gamma\beta'\gamma'} =\bar S_{\alpha\beta}{\bar{S'}}_{\alpha'\beta'} \Big( (8{x'}_{\alpha'} + 2\bar\partial'_{\alpha'})(x'\cdot\theta_{\alpha})+2{x'}_{\alpha'}(\theta_{\alpha}\cdot\bar\partial')\Big){\cal{W}}^H_{\beta\gamma\beta'\gamma'},$$
$$ {Q'}_{0}^{(1)} {\cal{W}}^H_{\beta\gamma\beta'\gamma'} =0.\;\;\;\;\;\;\;\;\;\;\;\;\;\;\;\;\;\;\;\;$$

As stated so far, it is the work of a few lines to show that
\begin{equation} \begin{aligned} \label{wituout'}
{\cal W}_{\alpha\beta\gamma\alpha'\beta'\gamma'}(x,x')&={\bar{S}}_{\alpha \beta}{\bar{S}'}_{\alpha' \beta'}\Big[ (\bar\partial'_{\alpha'} + {x'}_{\alpha'}) \Big( -\frac{1}{2} (x'\cdot\theta_{\alpha}) + \frac{1}{8} (\theta_{\alpha}\cdot\bar\partial') \Big) + (\theta' _{\alpha'}\cdot \theta_{\alpha})\Big]{\cal{W}^H_{\beta\gamma\beta'\gamma'}} \\ &-\frac{1}{8} {\bar{S}}_{\alpha \beta}{\bar{S}'}_{\alpha' \beta'} (\bar\partial'_{\alpha'} + {x'}_{\alpha'}) \Big({x'}_{\beta'} \theta_{\alpha}\cdot {\cal{W}^H_{\beta\gamma\cdot\gamma'}} + {x'}_{\gamma'}\theta_{\alpha}\cdot {\cal{W}^H_{\beta\gamma\beta'\cdot}}\Big),
\end{aligned}\end{equation}
where $ {\cal{W}^H_{\beta\gamma\beta'\gamma'}}$ is \cite{dehghani}
\begin{equation} \begin{aligned}
{\cal{W}^H}(x,x')= \Big(-\frac{2}{3} S\theta'\theta\cdot + S'S\theta'\cdot\theta &\\ + \frac{1}{3}S'S(\bar\partial' - x')[x'\cdot\theta& + \frac{1}{9}\bar\partial'\theta\cdot] \Big) \Big(\theta'\cdot\theta - \frac{1}{2}\bar\partial'[\theta\cdot\bar\partial' + 2\theta\cdot x'] \Big){\cal W}_{mc}(x,x').
\end{aligned} \end{equation}

Meanwhile, the dS minimally coupled massless scalar field two-point function, ${\cal W}_{mc}$, has been found in \cite{folacci} as follows
\begin{equation} {\cal W}_{mc}(x,x') = \frac{1}{8\pi^2}\left[\frac{1}{1-{\cal{Z}}(x,x')}-\ln(1-{\cal{Z}}(x,x'))+\ln 2+f(\eta,\eta')\right],\end{equation}
it is worth to mention that, $ {\cal{Z}}$ is an invariant object under the isometry group $O(1,4)$ which is defined for two given points on the dS hyperboloid $x$ and $x'$, by
$$ {\cal{Z}}\equiv -x.x'=1+{\frac{1}{2}}(x-x')^2,$$
so that, any function of ${\cal Z}$ is dS-invariant, as well. Whereas, $f$ is a function of the conformal time $\eta$ that breaks the dS invariance. In addition, the term $ln(1-{\cal Z}(x,x'))$, at largely separated points, is responsible for the advent of the infrared divergence. However, by constructing a covariant quantization of the massless minimally coupled scalar field through Krein space quantization, we have \cite{gazeaurenaud, takook3}
\begin{equation} \label{twopointkrein} {\cal W}_{mc}^{Krein}(x,x')=\frac{i}{8\pi^2}\epsilon (x^0-x'^{0})\left[\delta(1-{\cal{Z}}(x,x'))+{\vartheta({\cal{Z}}(x,x')-1)}\right],\end{equation}
where ${\vartheta}$ is the Heaviside step function and
\begin{equation} \epsilon (x^0-x'^0)=
\left\{\begin{array}{rl} 1& \;\;\;\;\;\;x^0>x'^0 ,\vspace{2mm}\\
0& \;\;\;\;\;\;x^0=x'^0,\vspace{2mm}\\
-1& \;\;\;\;\;\; x^0<x'^0.
\end{array}\right.
\end{equation}
Note that this two-point function has been written in terms of ${\cal Z}$, thus the de Sitter invariance is indeed preserved. It is also free of any pathological large-distance behavior.

\section{Conclusion}
A group theoretical approach to quantum gravity, based on the Wigner's theorem and Dirac's six-cone formalism, led to the CI field equation for the massless spin-2 field in de Sitter space \cite{takook}. In this paper, the corresponding CI two-point function was computed. The calculations were carried out through Krein quantization method. This method has already been successfully applied to the massless minimally coupled scalar field in de Sitter space-time for which it preserves covariance \cite{gazeaurenaud, De Bievre}. On this basis, it was shown that the two-point function is dS invariant and also free of any infrared divergences.

At the end, we would like to mention that, although, the geometrical interpretation of this linear theory is not entirely clear, but it may have an interesting property linked to quantum approach to the modified gravitational theories, say metric-affine theories of gravity. The advent of a rank-3 tensor field implies that, contrary to General Relativity (GR) assumptions, the space-time geometry is not fully described by the metric only, and other geometrical objects which can be independent of metric, such as connections, must be taken into account. In general, the connection does carry dynamics, so that the theory presents more degrees of freedom than GR. Consequently, torsion\footnote{The antisymmetric part of the connection is often called the Cartan torsion tensor.}does not remain non-propagating \cite{f(R) gravity (1)}.

Actually, if we accept that quantum theory of gravity should be an effective field theory, as many do \cite{QG}, we can conclude remarkable results; It is proved that, torsion is zero in vacuum and in the presence of a scalar field or the electromagnetic field, however, in the presence of a Dirac field or other vector and tensor fields it does not necessarily vanish \cite{f(R) gravity (1)}. This shows a correspondence between torsion and the presence of fields that describe particles with spin. So, though when torsion is present, the concept of a perfect fluid has to be generalized if one wants to include particles with spin, but since many cosmological and astrophysical applications are related to either the vacuum or the environments where matter can more or less be accurately described as a perfect fluid, these contributions to torsion will be negligible in most cases \cite{f(R) gravity (2)}. Therefore, it seems that these dynamical degrees of freedom can be eliminated in low-energy regimes \cite{f(R) gravity (1)},\footnote{It is expected that at some intermediate or high energy regimes, the spin of particles might interact with the geometry \cite{f(R) gravity (3)}.} and still, one can consider the dS space-time as the classical background with good accuracy. Nevertheless, we believe that in high-energy physics, where quantum corrections are important, these effects cannot be ignored. In this respect, the calculated two-point function may have an important role in formulating the future theory of quantum gravity.

\setcounter{equation}{0}
\begin{appendix}
\section{Mathematical Relations Underlying the Eq. (\ref{2.14-1})}
Regarding the Eqs. (\ref{2casimir}) and (\ref{S and M}), the action of the Casimir operator $Q^{(1)}$ on a rank-3 tensor field can be written as follows
\begin{equation} \label{casimir}\begin{aligned}
Q^{(1)} F_{\alpha\beta\gamma}& = (Q_{0}^{(1)}-6)F_{\alpha\beta\gamma} + 2 \Big(\eta_{\alpha\beta}F_{\delta\delta\gamma} + \eta_{\beta\gamma}F_{\alpha\delta\delta} + \eta_{\alpha\gamma}F_{\delta\beta\delta} \Big) +2 \Big( x_\alpha\partial\cdot F_{\cdot\beta\gamma} + x_\beta\partial\cdot F_{\alpha\cdot\gamma}\\ + & x_\gamma\partial\cdot F_{\alpha\beta\cdot} \Big) - 2\Big( \partial_\alpha x\cdot F_{\cdot\beta\gamma} + \partial_\beta x\cdot F_{\alpha\cdot\gamma} + \partial_\gamma x\cdot F_{\alpha\beta\cdot} \Big) - 2\Big( F_{\beta\alpha\gamma} + F_{\gamma\beta\alpha} + F_{\alpha\gamma\beta} \Big).
\end{aligned}\end{equation}
It is important to note that by imposing the following conditions on the tensor field:
\begin{itemize}
\item $F_{\alpha\beta\gamma} = - F_{\beta\alpha\gamma}$ and $\sum_{cycl}F_{\alpha\beta\gamma} = F_{\alpha\beta\gamma}+ F_{\beta\gamma\alpha} + F_{\gamma\alpha\beta} =0 $; the mixed symmetry conditions. Note that, these conditions are necessary for UIRs of the conformal group \cite{binegar}.
\item $ x\cdot F_{\cdot\beta\gamma} = x\cdot F_{\alpha\cdot\gamma} = x\cdot F_{\alpha\beta\cdot} =0 $; the transversality conditions.
\item $ \partial\cdot F_{\cdot\beta\gamma} = \partial\cdot F_{\alpha\cdot\gamma} = \partial\cdot F_{\alpha\beta\cdot} =0 $; the divergenceless conditions. Note that, for transverse tensors, like $F_{\alpha\beta\gamma}$, $ \partial\cdot F_{\cdot\beta\gamma} = \bar\partial\cdot F_{\cdot\beta\gamma}$.
\item $ F_{\alpha\delta\delta} = 0 $; the traceless condition.
\end{itemize}
which are necessary for UIRs of the dS and conformal groups, the Eq. (\ref{casimir}) reduces to
$$ (Q^{(1)} + 6) F_{\alpha\beta\gamma} = Q_{0}^{(1)} F_{\alpha\beta\gamma}. $$
For more mathematical details of the action of the Casimir operators ($Q^{(1)}$ and $Q^{(2)}$), the commutation rules and algebraic identities of the various operators and fields, one can refer to \cite{dehghani,pejhan}

\setcounter{equation}{0}
\section{Mathematical Relations Underlying the Eq. (\ref{kh})}
Generally, the following form for $K_{\beta\gamma}$ can be considered
\begin{equation} \label{general}\begin{aligned}
K_{\beta\gamma} = \; &C_1 (x\cdot Z H_{\beta\gamma}) + C_2 (Z\cdot\bar\partial H_{\beta\gamma}) + C_3 (\bar\partial_\beta Z\cdot H_{\cdot\gamma} + \bar\partial_\gamma Z\cdot H_{\beta\cdot}) + C_4 (x_\beta Z\cdot H_{\cdot\gamma}\\ &+ x_\gamma Z\cdot H_{\beta\cdot})
+ C_5 (\theta_{\beta\gamma}Z\cdot H \cdot Z) + C_6 (\bar\partial_\beta\bar\partial_\gamma - x_\gamma\bar\partial_\beta)Z\cdot H \cdot Z,
\end{aligned}\end{equation}
clearly $ K_{\beta\gamma} = K_{\gamma\beta} $. $C_1$, ... $C_6$ are six arbitrary real numbers, which are determined by considering the following physical requirements: \\
The transversality conditions ($x\cdot K_{\cdot\gamma}=x\cdot K_{\beta\cdot}=0$) require that
\begin{equation} \label{transversality conditions} C_2 + C_3 +C_4=0.\end{equation}
And then the condition (\ref{conditions}$-III$) makes
\begin{equation} \label{trace} C_5 = - C_6,\;\; \mbox{and} \;\; C_1 + 4C_4 + 1=0.\end{equation}
Regarding the conditions (\ref{conditions}$-I$ and $II$), one can obtain
\begin{equation} \label{div} C_1 = -\frac{1}{2},\;\; C_4=-\frac{1}{8},\end{equation}
and also a new auxiliary equation $ \bar\partial_\beta Z\cdot H_{\cdot\gamma} = x_\gamma Z\cdot H_{\cdot\beta}$, which states that the third and fourth terms in (\ref{general}) are not independent, so, without any damage to the generality of the solution, one can take $C_3 = 0$. Then we have $ C_2= \frac{1}{8} $, and so, one can rewrite the general solution for $K_{\beta\gamma}$ as follows
\begin{equation}\label{final} K_{\beta\gamma} = \; -\frac{1}{2} (x\cdot Z H_{\beta\gamma}) + \frac{1}{8} (Z\cdot\bar\partial H_{\beta\gamma}) - \frac{1}{8} (x_\beta Z\cdot H_{\cdot\gamma}+ x_\gamma Z\cdot H_{\beta\cdot})
+ C_5 ( \bar\partial_\beta x_\gamma -\bar\partial_\beta\bar\partial_\gamma )Z\cdot H \cdot Z.\end{equation}
Note that, a straightforward calculation shows that the Eq. (\ref{ar}) does not create new constraints to be imposed on (\ref{final}). Therefore, since we're looking for the easiest possible answer, we choose $C_5 = 0 $.
\end{appendix}

\end{document}